# PROBABLE IDENTIFICATION OF THE IMPACT CRATERS ASSOCIATED WITH TWO LUMINOUS HISTORICAL LUNAR FLASHES

William Bruckman (miguelwillia.bruckman@upr.edu ) and Abraham Ruiz[1] (abraham.ruiz@upr.edu)

[1] University of Puerto Rico at Humacao, Department of Physics and Electronics

Abstract: We have reviewed and analyzed the lunar luminous events observations by William Herschel during the peak of the Lyrid meteor shower of 1787 and Leon Stuart near the peak of the Leonid meteor shower of 1953, seeking the impact craters that these events presumably formed. Evidence is presented that identifies two cold spot fresh craters as the expected candidates.

1 INTRODUCTION

The development of direct observations of meteorite impacts on the moon, called Lunar Flashes, has given rise to an important field of observational determination of the impact rate at the lower range of meteorite energies (see for instance [1] for an overview and references on lunar flashes observations in the last 30 years). In what follows, we will discuss two historical observations of luminous lunar events, associated with one Lyrid and one Leonid meteor showers, that were perhaps the first scientifically well documented lunar flashes, and we will also present evidence for the probable identification of the corresponding two craters they formed. One of them was the observations by William Herschel of three luminous sources on the lunar surface, on April 19 and 20, 1787, at the peak of the Lyrid meteor shower [2], that he interpreted as being produced by three volcanoes. The other, referred to as the Stuart flash event, occurred on November 15, 1953, near the peak of the Leonid meteor shower [3], and was photographed by Leon H. Stuart, a physician and amateur astronomer who serendipitously observed the lunar flash while preparing to take astronomical telescopic pictures. A present-day interpretation of these events is that they were the result of meteorite impacts associated with the Lyrid meteorite shower [4] and the Leonid shower.

2 THE WILLIAM HERSCHEL EVENT

Below, we will review some of the evidence presented in reference [4] for the interpretation of Herschel observations as caused by a meteorite that produced impact melt and the fresh crater of approximately 750 meters in diameter, illustrated in Figures 1 and 2.

Here is part of William Herschel's description of the event [2]:
"April 19, 10h. 36` sidereal time: I perceive three volcanoes in different places of the dark part of the new moon. Two of them are either already nearly extinct, or otherwise in a state of going to break out; which perhaps may be decided next lunation. The third shows an actual eruption of fire, or luminous matter. I measured the distance of the crater from the northern limb of the moon, and found it to be 3` 57".3. Its light is much brighter than the nucleus of the comet which M. Mechain discovered at Paris the 10[th] of this month; April 20, 10h. 2` sidereal time: The volcano burns with greater violence than last night. I believe its diameter cannot be less than 3``, by comparing it with that of the Georgian planet; as Jupiter was near at hand, I turn the telescope to his third satellite, and estimated the diameter of the burning part of the volcano to be equal to at least twice that of the satellite. Hence we may compute that the shining or burning matter must be above three miles in diameter. It is of an irregular round figure, and very sharply defined on the edges …. The appearance of what I have called the actual fire of eruption of a volcano, exactly resembled a small piece of burning charcoal, when it is covered by a very thin coat of white ashes, which frequently adhere to it when it has been some time ignited; and it had a degree of brightness, about as strong as that with which such a coal would be seen to glow in faint daylight …."

Herschel indicated that the actual eruption of luminous matter was located at an angular distance of 3` 57".3 from the northern limb of the moon. At approximately this position, we found a cold spot crater with a diameter of about 750 m (Figures 1 and 2). Cold spots (Bandfield et al. [5]) constitute a family of fresh lunar craters that are characterized by having a surface surrounding the impact with a nighttime average temperature at least 2 kelvin degrees less than the background rock free regolith temperature (shown in shades of blue in Figures 1 and 2). This property seems common to all new and small impacts (100 m $\approx \leq D \leq \approx$ 1.5 km), but ephemeral, so that the total number of them is only about 4,000 [6], and therefore they are expected to be relatively recently formed. On the other hand, the area interior to the above candidate crater and its immediate surroundings, up to a few diameters, has higher than average temperatures (shown in shades of red in Figures 1 and 2), which is consistent with the presence of boulders in this area. This temperature duality happens when the cold spot craters are large enough. Boulders and warmer interior areas could also be associated with melt events [6], and such impact melts could perhaps explain the description by Herschel of a prolonged visibility for more than a night with the following characteristics [2]:

"The appearance of what I have called the actual fire of eruption of a volcano, exactly resembled a small piece of burning charcoal, when it is covered by a very thin coat of white ashes, which frequently adhere to it when it has been some time ignited; and it had a degree of brightness, about as strong as that with which such a coal would be seen to glow in faint daylight …."

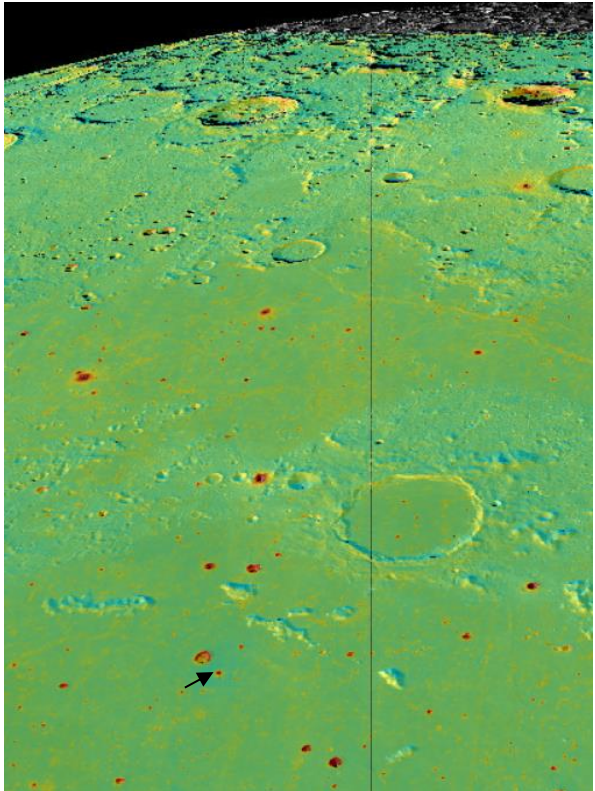

Figure 1: The cold spot (arrow) at latitude 45.985 and longitude –14.715 degrees. LROC Diviner view. Blue colors denote lower temperatures and red colors higher ones.

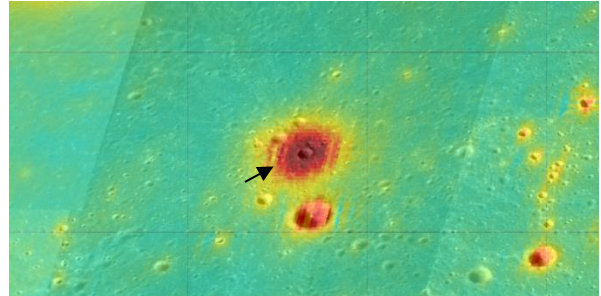

Figure 2: Amplification of the cold spot of Figure 1.

3 THE STUART FLASH

Another well documented bright Lunar flash is the well known Stuart flash, which occurred on November 15, 1953, near the peak of the Leonid meteor shower. It was photographed by Leon H. Stuart [3], who observed the flash as he was taking pictures of the moon. We have investigated the lunar location corresponding to the luminous area associated with the Stuart flash, searching, with the use of LROC images, for the expected youthful-looking impact crater formed by the flash. We found a candidate cold spot crater with a diameter of approximately 160 m near the northwest border of the flash area, about 38 km from the center of the flash (Figure 3, blue dot). This crater may not look like a good candidate for the source of the Stuart event, since is not at the center of the Stuart flash bright area, which is estimated to be located by the orange dot in Figure3. See: Mapping Lunar Impact flashes, https://meetingorganizer.copernicus.org/EPSC2017/EPSC2017-971.pdf ; however, in the next paragraph we will elaborate that the flash center need not coincide with the impact point.

There is direct observational evidence that a cloud of luminous material generated by an impact flash moved tens of kilometers, reaching a considerable distance from the original impact [8,9]. The photographs, video, and analysis of this impressive impact event on February 26, 1915 (see https://hobbyspace.com/Blog/?p=10165) allow us to conclude that a similar dynamic for the transportation of the cloud associated with the cold spot impact crater under consideration here could be responsible for what was photographed by Stuart in the flash event in 1953.

A further evidence is an intriguing observation, using the LROC WAC Basemaps $TiO_2$ Abundance, that shows a path (see Figure 4) of excess $TiO_2$ material from the impact site of the cold spot crater to approximately the center of the Stuart event flash. This could be interpreted as ejected material transported from a region richer in $TiO_2$ toward a sector with a low percentage of $TiO_2$.

Also, it is interesting that an inspection of the cold spot crater in Figures 5 and 6 reveals elongated lobe-shaped features in the direction of the Stuart flash center. They appear to have resulted from an impact at an inclined angle that produced ejected material in this direction.

 4 CONCLUSIONS

We have presented evidence that the two cold spot craters in Figures 2 and 6 were the consequences of the meteorites likely responsible for the historical lunar observations by William Herschel in 1787 and Leon Stuart in 1953, respectively.

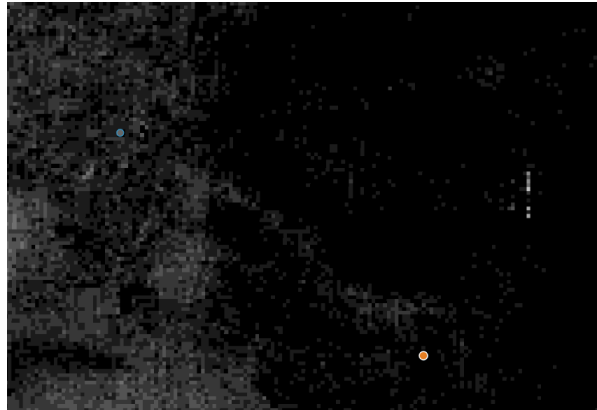

Figure 4: Note the path of a higher weight percent of $TiO_2$ in white color approximately connecting the center of the cold spot crater (blue circle) with the center of the Stuart flash (orange circle). LROC WAC Basemaps, $TiO_2$ abundance.

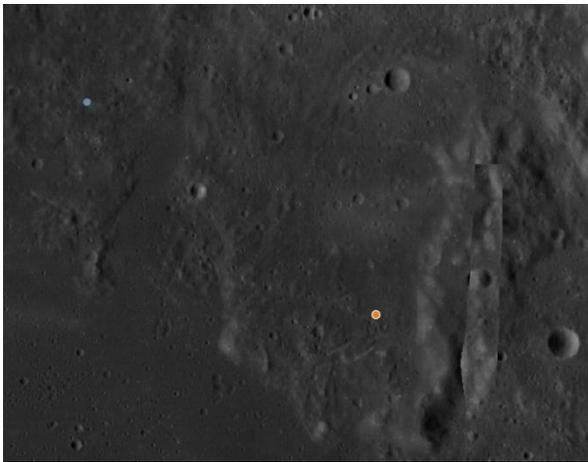

Figure 3: Location of Stuart flash center according to reference [7]: https://meetingorganizer.copernicus.org/EPSC2017/EPSC2017-971.pdf (orange circle); latitude 4.29, longitude –3.30 degrees. Also, location of cold spot crater center (blue circle); latitude 5.053, longitude –4.32 degrees.

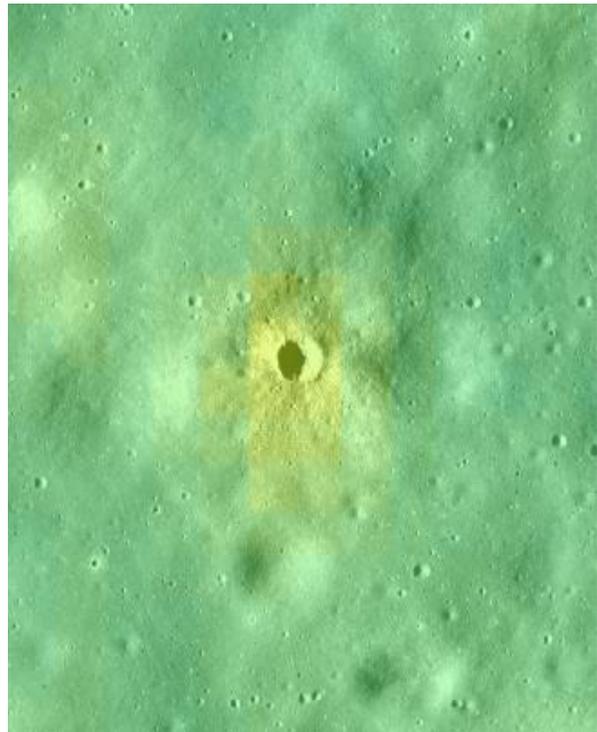

Figure 5: The cold spot crater with coordinates of latitude 5.053 and longitude –4.33 degrees, at the center of the figure, that we are associating with the Stuart flash event.

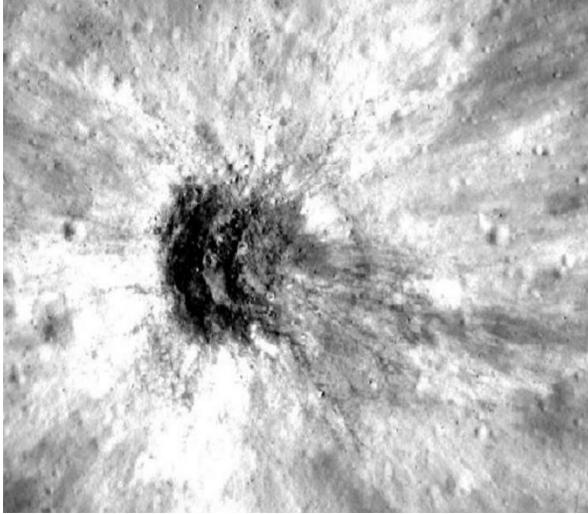

Figure 6: The crater in Figure 5 at small solar inclination, showing an asymmetric shape with ejected material toward the southeast.